# Metasurface-enhanced mid-infrared spectrochemical imaging of tissues

S. Rosas[1], K. A. Schoeller [3,4], E. Chang[3,4], H. Mei[2], M.A. Kats[2], K.W. Eliceiri[1], X. Zhao[3,4],

F. Yesilkoy[1, *]

[1]Department of Biomedical Engineering, University of Wisconsin-Madison, Madison, WI, USA

[2]Department of Electrical and Computer Engineering, University of Wisconsin-Madison, Madison, WI, USA

[3]Department of Neuroscience, University of Wisconsin-Madison, Madison, WI, USA

[4]Waisman Center, University of Wisconsin-Madison, Madison, WI, USA

*Corresponding author: filiz.yesilkoy@wisc.edu

## Abstract

Label-free and nondestructive mid-infrared vibrational hyperspectral imaging is emerging as an important ex-vivo tissue analysis tool, providing spatially resolved biochemical information critical to understanding physiological and pathological processes. However, the chemically complex and spatially heterogeneous composition of tissue specimens and the inherently weak interaction of infrared light with biomolecules limit the analytical performance of infrared absorption spectroscopy. Here, we introduce an advanced mid-infrared spectrochemical tissue imaging modality using metasurfaces that support strong surface-localized electromagnetic fields to capture quantitative molecular maps of large-area murine brain-tissue sections. Our approach leverages polarization-multiplexed multi-resonance plasmonic metasurfaces to simultaneously detect many different functional biomolecules. The resulting surface-enhanced mid-infrared spectral imaging (SE-MIRSI) method eliminates the non-specific effects of bulk tissue morphology on the quantitative analysis of fingerprint spectra and improves the chemical selectivity. We show that the metasurface enhancement increases the retrieval of amide I and II absorption bands associated with secondary structures of proteins. Moreover, we demonstrate that plasmonic metasurfaces enhance the chemical contrast in infrared images and enable the detection of ultrathin tissue regions that are not otherwise visible to conventional mid-infrared spectral imaging. While we tested our approach on murine brain tissue sections, this chemical imaging method is well-suited for any tissue type, which significantly broadens the potential impacts of our method for both translational research and clinical histopathology.

*In vitro* tissue imaging is critical for understanding many physiological and pathological processes relevant to biomedical research and clinical diagnostics. In particular, hyperspectral imaging modalities are emerging across the electromagnetic spectrum, combining the strengths of high-resolution spectral and spatial information to generate new contrast mechanisms for diverse morphological and biochemical properties of tissues. The mid-infrared (MIR) spectrum is uniquely attractive for hyperspectral imaging of tissues because of the abundance of vibrational resonances that form the fingerprints of functional biomolecules.[1,2] Because it is nondestructive and label-free, MIR spectroscopy is also well-suited for correlated multimodal tissue imaging and can provide essential complementary biochemical information to the current gold-standard modalities, including immunohistochemistry,[3] proteomics by mass spectrometry,[4] and gene expression profiling.[5] Moreover, the recent remarkable advancements in infrared photonics, including the development of high-power quantum cascade lasers (QCLs)[6] and high-performance infrared detector arrays, have empowered the development of MIR hyperspectral imaging instruments.[7,8] Notably, when the high-content hyperspectral tissue image datasets are processed using machine learning (ML) classifiers and other artificial intelligence, [9,10,11] they can bring new insights into biomedical sciences[12] and measurable advances in clinical diagnostics.[13]

Despite the potential of mid-infrared spectral imaging (MIRSI) for histology, current approaches fall short in quantitative chemical analysis of complex and heterogeneous tissue specimens. The fundamental limitation of standard infrared absorption spectroscopy is that it yields intermingled chemical and morphological information about the specimen. Specifically, Beer's Law is typically used for quantitative analysis in MIRSI: $A(v) = \sum_i^N [a_i(v)c_i l]$, where the total measured absorption signal ($A$) at a specific wavenumber ($v$) depends on the absorptivity ($a_i(v)$) of N different species, their concentrations ($c_i$), and optical path length ($l$). When working with complex and spatially heterogeneous samples, variations in $A(v)$ can be attributed to changes in any one or more of the three parameters ($a_i, c_i, l$). Therefore, the spectral information acquired from a specific spatial position is usually congested by the intense bands of the abundant molecules in complex specimens. For instance, MIR spectroscopy is an excellent analytical tool for the conformational analysis of proteins through the sensitive amide II & I region, which includes a variety of vibrational resonances corresponding to the secondary backbone structures of proteins.[14] Such structure analysis is important for biomedical investigations of neurodegenerative diseases.[15] However, identification of individual bands associated with protein backbone structures, such as α, and β-sheet, coils, and turns of amide bands, from complex tissue samples is challenging due to the bulk tissue absorption and scattering artifacts contributing to high baselines.[16,17,18] Thus, to unleash the full potential of MIRSI as an effective chemical tissue imaging method, there is an unmet need for new measurement modalities to mitigate chemical vs. morphological ambiguity and address spectral congestion in heterogeneous samples.[19]

Recently, harnessing light via nanoengineered materials has enabled light confinement beyond the diffraction limit of conventional optical imaging.[20,21,22] Specifically, resonant electromagnetic metasurfaces can generate exceptional field confinement within ultrathin volumes and thus can enhance light-matter interactions in optical spectroscopies. Previously, resonant metasurfaces have been used for surface-enhanced infrared absorption spectroscopy (SEIRAS), where plasmonic nanoantennas with resonances in the MIR have been used to push the limits of efficiency in light-matter interactions.[23] SEIRAS has been applied to detect protein monolayers,[24] lipid bilayers,[25] thin polymer films,[26] self-assembled molecular monolayers,[27] and to monitor and classify biological cells. [28,29] However, SEIRAS has not yet been used for unambiguous chemical imaging of complex real-world tissue samples to reveal their spatially heterogeneous chemical composition.

Here, we present metasurface-enhanced MIR spectral imaging (SE-MIRSI) as an advanced label-free chemical tissue-imaging method (**Figure 1**). We demonstrate MIR spectroscopy of murine brain tissue sections, interrogating nanoscale thicknesses over large areas using engineered photonic substrates and a hyperspectral imaging microscope with illumination from four tunable QCLs (Figure 1a). To satisfy the critical requirement of multiplexed molecular fingerprint analysis of SE-MIRSI, we developed polarization-tunable multi-resonant plasmonic metasurfaces that enhance light-matter interactions in sub-wavelength volumes over our target spectral range of 950–1800 $cm^{-1}$ (Figure 1b and 1c). We reveal how tissue morphology affects quantitative chemical analysis by comparing absorbance from numerous vibrational bands, measured from murine brain tissue sections of different thicknesses, ranging from 20 μm to 80 μm, mounted on standard $CaF_2$ substrates or plasmonic metasurfaces. We show that SE-MIRSI is not affected by bulk morphological properties of complex tissue samples because the measured spectra exclusively capture light-matter interactions at the metasurface-generated plasmonic hotspots – nanoscale volumes with strong enhancement of the electric field intensity (~$1.4\times10^3$). Thus, SE-MIRSI decouples the molecular composition information from the physical properties of the sample, such as tissue thickness and morphology. We also show that SE-MIRSI can detect fine spectral details corresponding to protein backbone structures, as well as other important biomolecules, such as glycogen, nucleic acids, proteins, and lipids (Figures 1d and 1e). Finally, we show that SE-MIRSI enables chemical analysis of ultrathin brain tissue regions, which are not otherwise detectable on standard substrates, underscoring the high analytical sensitivity of our approach. Our method, *for the first time*, addresses a well-known quantitative analysis limitation of infrared absorption spectroscopy by decoupling chemical and morphological properties and enabling highly sensitive and chemically selective analysis of complex and heterogeneous biosamples. Our chemical imaging method is versatile and can be used to study any tissue type with potential applications in fundamental research and clinical histopathology.

## Results

To elucidate the quantitative and qualitative effects of physical tissue thickness variations on spectral measurements, we characterized adjacent sections of murine brains of three different thicknesses (20, 40, and 80 μm). Coronal brain tissue sections were obtained from an 8-weeks-old (young adult) wild-type male mouse (details of tissue preparation protocol in Methods). We mounted the adjacent brain sections on standard MIR-transparent $CaF_2$ substrates and collected hyperspectral image data of the hippocampal region using conventional MIRSI in transmission mode (**Figure 2a**). Each MIRSI measurement includes 480 × 480-pixel images acquired between 950-1800 $cm^{-1}$, with 2 $cm^{-1}$ spectral resolution. Our MIR microscope can achieve down to ~5 μm diffraction-limited spatial resolution, capturing a 650 × 650 $μm^2$ field-of-view (FoV) in less than a minute. Figure 2b shows representative single-pixel spectra from the three tissue sections. Herein, we specifically plot spectra from identical lateral regions of the adjacent brain tissue sections to minimize compositional differences. As expected, in transmission mode using conventional substrates, the characteristic absorption consistently increases with tissue thickness. For example, for the 40 μm thick tissue section, the amplitude of the amide I absorption band increased 2-fold with respect to the 20 μm thick tissue sample, following Beer's Law (Figure 2b). Moreover, we found that the tissue thickness affects the total absorbance signal and changes fingerprint spectra qualitatively, specifically in the strong amide I band (**Figure 2c** bottom). A possible explanation for this might be that some intense bands saturate, and other weaker bands become detectable as the tissue thickness increases.

For quantitative spectral analysis, we used derivative spectrometry.[30] Specifically, we used the second-order derivative of the measured absorbance spectra, which holds a linear relationship between amplitude ($A$) and concentration ($c_i$) (**Figure S1**). Figure 2c shows the second-derivative absorption spectra of the three tissue thicknesses in two separate plots to easily visualize the bands. An absorbance peak corresponds to a trough in the second-order derivative spectra. For quantitative analysis, we measured the area above these troughs by integrating the second-order derivative spectra over each band (Methods). Figure 2d shows the spatial pixel distributions ($n_{pixels}$ = 12.5×10$^3$) of the different bands' integrals from similar regions of adjacent tissue sections. As expected, the total absorbance data consistently show quantitative dependence on the tissue thickness on all the bands. Moreover, we identified tissue-thickness-dependent qualitative variations in the amide I band (Figure 2c bottom). Specifically, in the amide I region (1600 – 1700 cm$^{-1}$), the second-derivative spectrum from the 80 μm-thick tissue section has a double-trough spectral line shape compared to the single-trough in 20 and 40 μm-thick tissue sections. This underscores the qualitative fingerprint variations caused by the tissue morphology since we compare data collected from identical regions of adjacent brain tissue sections, expecting minimal variation in their chemical compositions. Together, this analysis highlights how in conventional MIRSI physical tissue properties can hinder access to accurate chemometric information reinforcing the need for the SE-MIRSI method.

Chemical analysis of complex and heterogeneous tissue samples using SE-MIRSI requires engineered substrates that can enhance light-matter interactions over a broad spectral range using spatially collocated and spectrally distinct strong resonance modes. Therefore, plasmonic structures with single-band resonances[24] or dielectric metasurfaces[31] supporting isolated sharp resonances are not compatible with the SE-MIRSI method. To address the requirements of the SE-MIRSI, we designed and fabricated a polarization-multiplexed multi-resonant broadband plasmonic metasurface, where each metaunit consists of a cluster of gold nanostructures in the shape of "*ulu*". The *ulu* design supports multiple strong plasmonic resonances with both orthogonal polarizations of the incident light (Figure 3b).[32] Herein, we use well-understood plasmonic structures as building blocks – a rod antenna[24] and two u-shaped split ring resonators (SRR)[33]. When excited with linearly polarized infrared radiation at a 90° polarization angle, the strong coupling between the elementary radiative electric dipolar resonance of the center rod and the non-radiative fundamental magnetic modes of the side SRRs leads to a mode hybridization[34] and results in spatially overlapping yet spectrally split plasmonic resonances at 1300 cm$^{-1}$ and 1900 cm$^{-1}$ (**Figure S2**). In the horizontal 0° polarization, the radiative magnetic mode of the SRRs are excited at 1600 cm$^{-1}$, which fills the spectral gap between the hybridized modes. The three strong plasmonic resonances densely cover our target fingerprint range, providing strong field enhancement to a broad set of vibrational resonance bands. **Figure 3a** shows a fabricated metasurface (5.5 × 5.5 mm$^2$) on a CaF$_2$ substrate using electron-beam lithography (Methods and Figure S7). The spectral position of each resonance peak is tuned to match target absorption bands by tailoring the geometric design parameters (Figures 3c and 3d). The details on physics of the plasmonic modes and their geometric tunability are given in SI (**Figures S2**, **S3**, **S4**).

In this work, we focus on eight absorption bands of interest associated with functional groups of glycogen (1024 cm$^{-1}$ ▲), deoxyribose (1060 cm$^{-1}$ ▲), symmetric phosphate (1090 cm$^{-1}$ ▲), C-OH (1170 cm$^{-1}$ ▲), asymmetric phosphate (1236 cm$^{-1}$ ▲), β-sheet (1542 cm$^{-1}$ ▲), Amide II (1574 cm$^{-1}$ ▲), and amide I (1660 cm$^{-1}$ ▲). Upon deposition of tissue sections on metasurfaces, the effective refractive index on top of the metasurface increases, leading to a spectral shift of the plasmonic resonance peaks towards lower wavenumbers. To compensate for this redshift, we designed the *ulu* structures to resonate at higher (by ~ 250 cm$^{-1}$) wavenumbers in the air (Figure 3c and 3d). Once the tissue section is mounted on the metasurface, spectrally overlapping vibrational resonances of molecules and plasmonic resonances couple to each other, generating plasmonic resonance damping at the wavenumbers associated with the vibrational bands. To simulate the effect of molecular absorbance on the plasmonic resonances, we used a dispersive medium described by the Lorentz model corresponding to each vibrational mode. For the 0° polarization, asymmetric phosphate PO$_2^-$ (1236 cm$^{-1}$ ▲) with a stretching vibration mode at 1093 cm$^{-1}$ was considered. For the 90° polarization, deoxyribose (1060 cm$^{-1}$ ▲), β-sheet (1542 cm$^{-1}$ ▲), amide II (1574 cm$^{-1}$ ▲), and amide I (1660 cm$^{-1}$ ▲) modes were simulated (Figures 3e and 3f). Importantly, the spatial overlap between the near-field hotspots generated by the three plasmonic resonances, such as the ends and edges of the rods in the *ulu* design, allows our SE-MIRSI method to analyze the same nanoscale sample volume via multiplexed quantitative detection of biomolecule signatures.

To show how SE-MIRSI addresses the quantitative chemical analysis limitations of conventional MIRSI in morphologically complex and heterogeneous tissue samples, we characterized two adjacent brain tissue sections with thicknesses 20 μm and 80 μm, deposited onto plasmonic metasurfaces (**Figure 4a**). Since we acquire hyperspectral image datasets in reflectance mode via the substrate side, the measured spectra exclusively capture light-matter interactions at the nanoscale metasurface-generated electromagnetic hotspot volumes but not the bulk interaction of light in the tissue. Our numerical findings indicate that the evanescent plasmonic field localized around the Au resonators decays exponentially, and surface enhancement in infrared spectra is only significant at a probing depth of ~100 nm (**Figure S5**). Similar to our MIRSI study above for quantitative band analysis, we consider the second-order derivative of the absorption spectra and integrals of the troughs associated with the target bands (Figures 4b and 4c). In Figure 4b, we compare the second derivative spectra collected by standard MIRSI and SE-MIRSI in four different spectral regions for 20 and 80 μm tissue thicknesses. Despite the significant difference in the spectral measurement modalities between MIRSI and SE-MIRSI, the detected bands show notable correspondence. Moreover, the SE-MIRSI spectral data from 20- and 80-μm-sections are similar and have features much like the MIRSI measurement for the 80-μm sample on CaF$_2$, which resolves weak amide I sub-bands (Figure 4b bottom right panel). Notably, with SE-MIRSI, we did not observe significant correlations between tissue thickness and total absorption signal computed via the integral of the second-order derivative spectra of each band (Figure 4c). We present further details on the statistical analysis of SE-MIRSI and MIRSI results comparing tissue thickness effects in the methods section and **Figure S9**. Our study shows that SE-MIRSI can decouple quantitative chemical information from tissue's physical thickness by confining light-matter interactions into nanoscale tissue sections using surface-localized fields to generate substantial absorbance signals from ultrathin volumes.

Due to the difficulties associated with the biochemical interpretation of individual vibrational bands in congested MIR spectra from complex biological samples, the spectral data are traditionally processed as a fingerprint and analyzed using bioinformatic tools.[19] Therefore, resolving vibrational resonances is critical for chemometric applications to enrich fingerprint data and increase chemical selectivity. To demonstrate the strong fingerprinting capabilities of the SE-MIRSI approach, we performed band assignment analyses on the SE-MIRSI spectra and compared our findings with standard MIRSI.[35] In **Figure 5a**, we show an example of a 20-µm-thick brain tissue section imaged in transmission mode at 1574 cm$^{-1}$ (amide II band ▲), showing areas on and off the metasurface, which is enclosed by white dashed lines. The resonant plasmonic metasurface visibly enhances molecular absorbance compared to tissue regions imaged on a standard substrate.

Moreover, we characterized the second-order derivative absorption spectra acquired by SE-MIRSI in reflection mode via the substrate side from four different brain regions, $R_1$, $R_2$, $R_3$, and $R_4$ on the metasurface (solid lines) (Figure 5b, 5c, and 5d). We also present a representative reference MIRSI measurement (dashed lines) of a brain tissue section in the same spectral range. First, our analysis reveals that spectral signatures collected by SE-MIRSI (Figure 5b, 5c, and 5d, solid lines) vary qualitatively and quantitatively in different brain regions. The evidence from this study suggests that SE-MIRSI can be used to perform chemical mapping of the brain regions. Second, discernible correspondence in band positions between the second-order derivative spectra collected by SE-MIRSI and MIRSI corroborates the reliability of our metasurface-enhanced spectral tissue imaging technique.

To further evaluate the fingerprint retrieval resolution of our approach, we perform individual band assignments on the absorption spectra from SE-MIRSI (Figure 5e, 5f, and 5g) and MIRSI (Figure 5h, 5i, and 5j). Here, we use the second-order derivative of the absorbance spectra to identify band-peak wavenumbers and then use them in band fitting and assignment processes (see Methods). Our results show distinct similarities in the vibrational bands we target in the same murine brain sample. Importantly, SE-MIRSI can capture some weak bands in the amide II & I spectral regions that conventional MIRSI does not detect. Therefore, our findings suggest that SE-MIRSI mitigates the congestion issue in conventional MIRSI, where strong bands from abundant chemical residues obscure the detection of weak and low-abundance bands due to bulk light-tissue interactions over long light paths. Moreover, in SE-MIRSI data, the amplitude of each band can be correlated to the abundance of their corresponding protein structures in every spatial pixel position enabling quantitative analytics of biosamples.

Next, we evaluate the resonance surface-enhancement effects by comparing the images acquired by MIRSI and SE-MIRSI approaches in an identical transmission-mode setup. **Figure 6a** shows single FoV images measured on a standard substrate (top row) and a metasurface illuminated at polarization angles of 0° (middle row) and 90° (bottom row). While for each FoV measurement, we collect 426 spectral images, here, we show eight functional groups' signature bands throughout the fingerprint spectrum. To facilitate comparison, we normalized absorbance over the three hyperspectral datasets with respect to the minimum and maximum absorbance values from the raw data. Thus, individual pixel values among the images and wavenumbers shown in Figure 6a can be compared. As designed, the plasmonic resonance excited at 0° polarization enhances absorption at C-OH and $PO^-_{2asym}$ bands, and the double-resonance peaks at 90° enhance glycogen, deoxyribose, $PO^-_{2sym}$, amide II, amide I, and ester bands. For example, the glycogen, deoxyribose, and $PO^-_{2asym}$ SE-MIRSI images at 90° polarization show significantly higher contrast than standard MIRSI and 0° polarization SE-MIRSI. To showcase the resonance effect further, we present histograms from $2.3 \times 10^5$ pixels on glycogen, C-OH, and $PO^-_{2asym}$, ester bands for three measurements (Figure 6b). Specifically, we focus on two different FoVs from the hippocampus region (Figure 6b, top row) and gray matter (Figure 6b, bottom row). The histogram data shows a consistent increase in overall pixel absorbance distributions when the target band overlaps with the plasmonic resonances at their corresponding wavenumbers. Due to the non-uniform porous morphology of the hippocampus region, the histogram at the top row shows a more corrugated distribution compared to the bottom row histograms showing data from a more uniform gray matter region.

In addition to enhancement in overall absorbance leading to increased contrast of spectral images, we further report that resonant SE-MIRSI enables chemical imaging of ultrathin tissue conformations that are not otherwise detectable. **Figure 7a** shows FoVs from hippocampus porous tissue regions imaged at 1730 cm$^{-1}$ corresponding to the C=O stretch of the ester band. We compared similar regions from two adjacent 20-μm-thick brain tissue sections on standard $CaF_2$ and our plasmonic metasurface. The lipid composition of ultrathin brain tissue residues is not detectable by conventional MIRSI nor off-resonance SE-MIRSI. In contrast, the presence of biomaterial is evidently visible on the resonant plasmonic metasurface due to the enhanced molecular detection sensitivity. To confirm that the contrast shown in images at 1730 cm$^{-1}$ is associated with the ester band, we plot second-order derivative spectra measured with the three modalities, i.e., on a standard substrate and the metasurface at 0° and 90° polarizations. These results further support that the resonant SE-MIRSI can enhance the analytical sensitivity and provide access to critical chemometric information from pixels that only reveal contrast when imaged on the resonant metasurfaces.

## Discussion

We have demonstrated three significant advancements to conventional MIRSI modality by employing strong light-matter interactions at the nanoscale sample volumes enabled by the engineered multi-resonant plasmonic metasurface. First, our experiments confirmed that by probing the complex tissue composition via metasurface-localized strong electromagnetic fields, we are able to decouple the effects of bulk physical sample parameters, i.e., tissue thickness and morphology, from the molecular composition information in spatially heterogeneous biological samples. Our approach provides a crucial solution to the well-known ambiguity of whether the total absorbance is related to the physical tissue morphology or analyte concentration in the MIR spectral data interpretation. It, thus, uniquely enables quantitative molecular mapping over large-area tissue sections.

Our second major finding is that SE-MIRSI is remarkably sensitive in resolving absorption bands of the fingerprint spectra. Specifically, we show multiplexed detection of functional molecular signatures in brain samples, including glycogen, nucleic acids, proteins, and lipids. Notably, SE-MIRSI can detect the amide I & II band peaks associated with the secondary backbone structures of proteins, which are usually challenging to resolve because of the abundant and strong bands that congest the spectra in bulk-tissue analyses. Thus, our innovative approach can be used to map brain regions for structural protein deformations to study molecular mechanisms in brain disorders. Finally, our novel ultrasensitive analytical approach can detect ultrathin tissue residuals based on their molecular content, which are invisible to conventional MIRSI.

Since the SE-MIRSI method leverages near-field interactions of the biosample with the plasmonic resonators, uniform adhesion of tissue sections to the metasurface chips is crucial for accurate measurements. To ensure that our tissue transfer procedure yields a uniform contact between the tissue and the metasurface, we measured a tissue sample transferred onto a metasurface chip before and after compressing the tissue between a glass slide and the metasurface chip. We did not observe a significant difference between spectral data from similar tissue regions with and without tissue compression (**Figure S6**). However, different tissue types prepared using alternative tissue preparation techniques may create adhesion issues. This can be addressed by coating the metasurface with a thin film of $SiO_2$, which was implemented before,[25] and then, modifying surface charges of the metasurface to ensure tissue adhesion to the substrate via ionic attractions.[36]

While developing the SE-MIRSI method, we collected data exclusively from fresh metasurface chips to eliminate any possible effects of chip recycling. However, chip recycling can be necessary once SE-MIRSI becomes an established tissue analysis technique to reduce substrate costs. A simple approach would be to conformally coat the metasurfaces with a thin film of $SiO_2$, MgO or $CaF_2$ using atomic later deposition or sputtering methods to introduce a thin protective layer. Then, the metasurfaces can be aggressively cleaned using harsh chemicals to enable their reusability. Additionally, wafer-scale fabrication methods were reported introducing opportunities to mass manufacture metasurfaces for a wide range of applications.[37,38] Thus, reusable metasurfaces fabricated using high-throughput technologies can enhance the accessibility of the proposed SE-MIRSI method enabling its broad applications in biomedical research and diagnostics.

We conclude that SE-MIRSI offers great promise as an ultra-sensitive, quantitative, rapid, and label-free chemical imaging technique for complex and heterogeneous tissue sample analyses. In this work, we tested our novel spectrochemical imaging modality on murine brain tissue sections because innovative imaging modalities are urgently needed to elucidate functional and molecular brain composition with high-throughput tools. Likewise, our approach can enable a host of exciting applications in the broad histopathology field, including biopsy tissue segmentations for diagnostics, prognostics, and personalized cancer therapies.

## Methods

**Electromagnetic simulations.** For the design of the metasurface, electromagnetic simulations were conducted using a finite-element frequency-domain solver on a tetrahedral mesh (with $1.3 \times 10^6$ total mesh cell) in CST Studio Suite 2022 with adaptive mesh refinement and periodic boundary conditions. The optical constants for the Au, Ti, and $CaF_2$ were taken from,[39],[40],[41] respectively.

**Nanofabrication of ulu plasmonic metasurfaces.** Double-side-polished, single-crystal $CaF_2$ (100) rounds (1 cm diameter) were used as substrates to fabricate the plasmonic metasurfaces. Nanostructures were fabricated by electron beam lithography, metal deposition, and a lift-off process sequence. Sequentially, two Poly (methyl methacrylate) (PMMA) resist layers were spin-coated at 4000 rpm and baked for 5 min at 180 °C. The bottom PMMA 495 A4 layer (~200 nm) helps with the lift-off process, and the top PMMA 950 A2 layer (~100 nm) was used as the high-resolution pattern-defining electron beam resist. A thin Cr layer (~10 nm) was sputtered on the PMMA to mitigate charging effects during electron beam exposure. Metasurfaces of 5.5×5.5 mm² area were patterned with a 100 keV electron beam. Following the resist exposure, the Cr layer was wet-etched, and then PMMA was developed in MIBK: IPA 1:3 solution for 90 sec. Au nanostructures were formed by electron beam evaporation of a ~5 nm Ti adhesion layer and a 90 nm-thick Au layer, followed by an overnight lift-off process carried out in acetone. To remove any PMMA residues from the plasmonic metasurfaces, an extra 2 min sonication in acetone and 1 min oxygen plasma treatment steps were necessary after the metal lift-off process (SI **Figure S7**).

**Mice brain tissue samples.** We performed all procedures involving live mice per the NIH Guide for the Care and Use of Laboratory Animals and the protocols approved by the University of Wisconsin-Madison Animal Care and Use Committee. Tissue processing and histological analyses of murine brain samples were performed as described previously with modifications.[42] Briefly, mice were euthanized by intraperitoneal injection of a mixture of 100 mg/kg of Ketamine, 20 mg/kg of xylazine, and 3 mg/kg of acepromazine, followed by transcardial perfusion with saline followed by 4% PFA. Brains were dissected, post-fixed overnight in 4% PFA. For chemical imaging, the brains were placed into cryomolds and submerged in optimal cutting temperature - O.C.T. compound on dry ice. Brain sections of 20 μm, 40 μm, and 80 μm thicknesses were cut using a sliding microtome and placed into DI water for 4 min to melt the surrounding O.C.T. Once the hippocampal region was reached, brain slices were mounted directly on previously unused plasmonic metasurfaces or $CaF_2$ rounds and stored at 4 °C until mid-infrared hyperspectral imaging measurements.

**Quantum-cascade-laser-based mid-infrared microscopy.** Mid-infrared spectral measurements were carried out using a quantum cascade laser (QCL) mid-infrared microscope. Spero-QT (Daylight Solutions, Inc.) is equipped with four QCL modules with spectral coverage from 950 to 1800 cm$^{-1}$. The microscope is continuously purged with dry nitrogen. A 4× IR objective (pixel size: 4.25 μm, 0.3 NA) and an uncooled microbolometer FPA with 480 × 480 pixels were applied for data collection in the reflection mode and a 12.5× IR objective (pixel size: 1.3 μm, 0.7 NA) in transmission mode. Data were collected with a 2 cm$^{-1}$ spectral resolution.

**Fourier-transform infrared spectroscopy measurements**. Transmittance and reflectance measurements were performed using a Fourier-transform infrared (FTIR) spectrometer coupled to an infrared microscope (Bruker Vertex 70 FTIR and Hyperion 2000). Reflectance and transmittance spectra were collected with linearly polarized light through a Casegrain reflective objective (15×, 0.4 NA) and measured by a liquid-nitrogen-cooled mercury-cadmium-tellurium (MCT) detector. Transmittance spectra of the plasmonic metasurface were calculated by normalizing the transmission spectrum by that of the $CaF_2$ substrate without nanostructures. Reflectance spectral measurements were referenced to a front-surface gold mirror.

**Statistical Analysis**

**Pre-processing of spectrometric data**. After hyperspectral 3D data collection, the data were preprocessed using Python functions and scripts. To filter out pixels from off-tissue substrate regions, we implemented a masking threshold based on the integral of the amide I band, from 1590 $cm^{-1}$ to 1720 $cm^{-1}$. The integral for the amide I range was calculated using the composite trapezoidal rule from Numpy (Figure 2b). For quantitative analyses, we used the second-order derivative of the absorption spectra to improve the accuracy of quantification (Figure 2c). For all spectral signals, the second-order derivative was calculated using Savitzky–Golay (SG) filter with a window length of 13 data points and a 2$^{nd}$ order polynomial (SG SciPy signal library in Python). In the SG filtering and derivative operations, 26 spectral data points out of 426 are lost in each pixel spectrum. Consistently, for quantitative analysis of each functional group, the area above the troughs was integrated using Equation (1) (Figure 2d). Reference integration limits are shown in **Table 1**. Finally, for the individual band fitting operations in Figure 5, gaussian band fitting was performed using non-linear least-squares minimization and Curve-Fitting (lmfit) in Python.

$$\text{Area above the trough} = \int_{Min}^{Max} \ddot{A}(\upsilon)\ d\upsilon \qquad (1)$$

Table 1 | Reference integration intervals along second-order derivative absorption spectra.

| Integration range ($cm^{-1}$) | Deoxy (1060 ▲) | $PO_{2sym}^-$ (1090 ▲) | $PO_{2asym}^-$ (1170 ▲) | $\beta$ (1542 ▲) | Amide II (1574 ▲) | $\alpha$ (1660 ▲) |
|---|---|---|---|---|---|---|
| Min | 1034 | 1074 | 1204 | 1522 | 1554 | 1610 |
| Max | 1078 | 1118 | 1270 | 1556 | 1584 | 1672 |

**Data presentation and statistical analysis.** All violin plots were built based on the spectrum of 12.5×10$^3$ pixels; the white dot represents the median; the thick grey bar in the center represents the interquartile range, and the shaded area represents the distribution of the integral values along each band. Wider and skinnier sections of the violin plots represent higher and lower probability, respectively.

We conducted a statistical analysis to demonstrate the effectiveness of our SE-MIRSI method in detecting and quantifying functional biomolecules from constant nanoscale tissue volumes compared to the conventional MIRSI. To accomplish this, we computed median absorbance values for 80 and 20 μm-thick tissue sections from the integral band distributions measured on the CaF$_2$ and the ulu-metasurfaces (data shown in Figures 2 and 4). Next, we calculated the ratio of the median absorbance values from 80 and 20 μm thick tissues for each band, which we call 80/20 ratio parameter. A two-tailed unpaired t-test was used to compare the 80/20 ratios between the brain tissue section on bare CaF$_2$ and on the plasmonic metasurface. A p-value of *p* = 5.74×10$^{-4}$ was identified between the 80/20 ratios for deoxyribose (1060 cm$^{-1}$ ▲), symmetric phosphate (1090 cm$^{-1}$ ▲), asymmetric phosphate (1236 cm$^{-1}$ ▲), and β-sheet (1542 cm$^{-1}$ ▲) bands (see **Figure S9**). In all cases, significance was defined as *p*<0.05.

## Acknowledgements

F.Y. acknowledges financial support from UW Carbone Cancer Center. K.W. Eliceiri acknowledges financial support from the National Institutes of Health (grant no. U54CA268069). M.A. Kats acknowledges financial support from the Office of Naval Research (grant no. N00014-20-1-2297). X. Z. acknowledges financial support from the National Institutes of Health (grant no. R01MH118827, R01MH116582, and R01NS105200 to X. Z., and grant no. P50HD105353 to the Waisman Center), Jenni and Kyle Professorship to X. Z. The authors thank Y. Xing in Zhao lab for technical assistance, Panksepp, D. Bolling, and MM Eastwood at the Waisman Animal Core for services.

## Disclosures

Authors declare no conflict of interest.

## Data availability

All data required and the code used to reproduce the results can be obtained from the corresponding author upon a reasonable request.

## Keywords

Plasmonic metasurface, mid-infrared spectral imaging (MIRSI), label-free brain tissue imaging, surface-enhanced infrared absorption spectroscopy (SEIRA), hyperspectral imaging, laser spectroscopy, nanobiophotonics, chemical imaging.

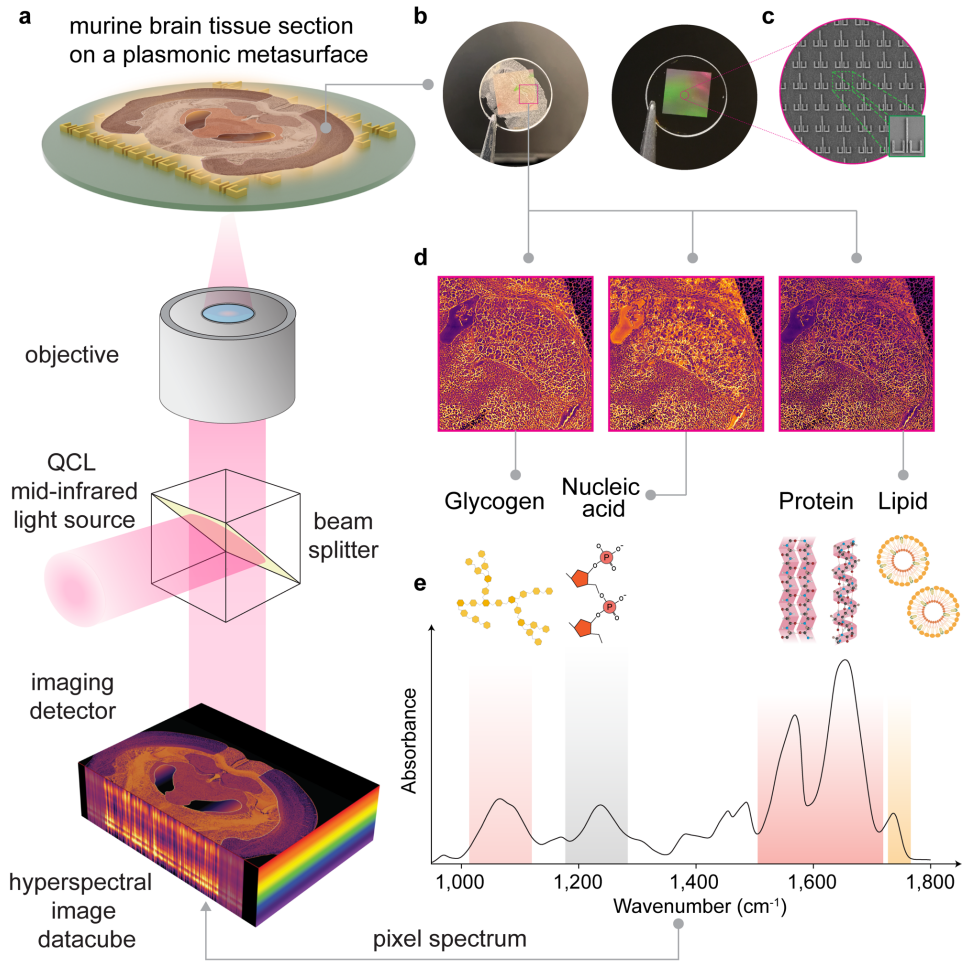

**Figure 1 | Surface-enhanced mid-infrared spectral imaging of tissues (SE-MIRSI). a**, Schematic of the tunable quantum cascade laser (QCL)-based mid-IR spectral imaging system. Hyperspectral datacubes are acquired using a microbolometer focal-plane detector array while sweeping the illumination wavenumber across the mid-IR fingerprint spectral range of 950 to 1,800 cm$^{-1}$. **b**, Photographs show a plasmonic metasurface chip fabricated over a 5.5 mm × 5.5 mm area on a CaF$_2$ substrate after and before mounting a murine brain tissue section. **c**, SEM images show an array of metaunits consisting of ulu-shaped Au nanostructures supporting polarization-multiplexed multi-resonance modes for broadband spectral coverage. **d**, SE-MIRSI of murine brain tissue section at different wavenumbers corresponding to absorption bands of functional groups present in endogenous biomolecules such as glycogens, nucleic acids, proteins, and lipids. **e**, A characteristic mid-IR absorption spectrum of mouse brain tissue.

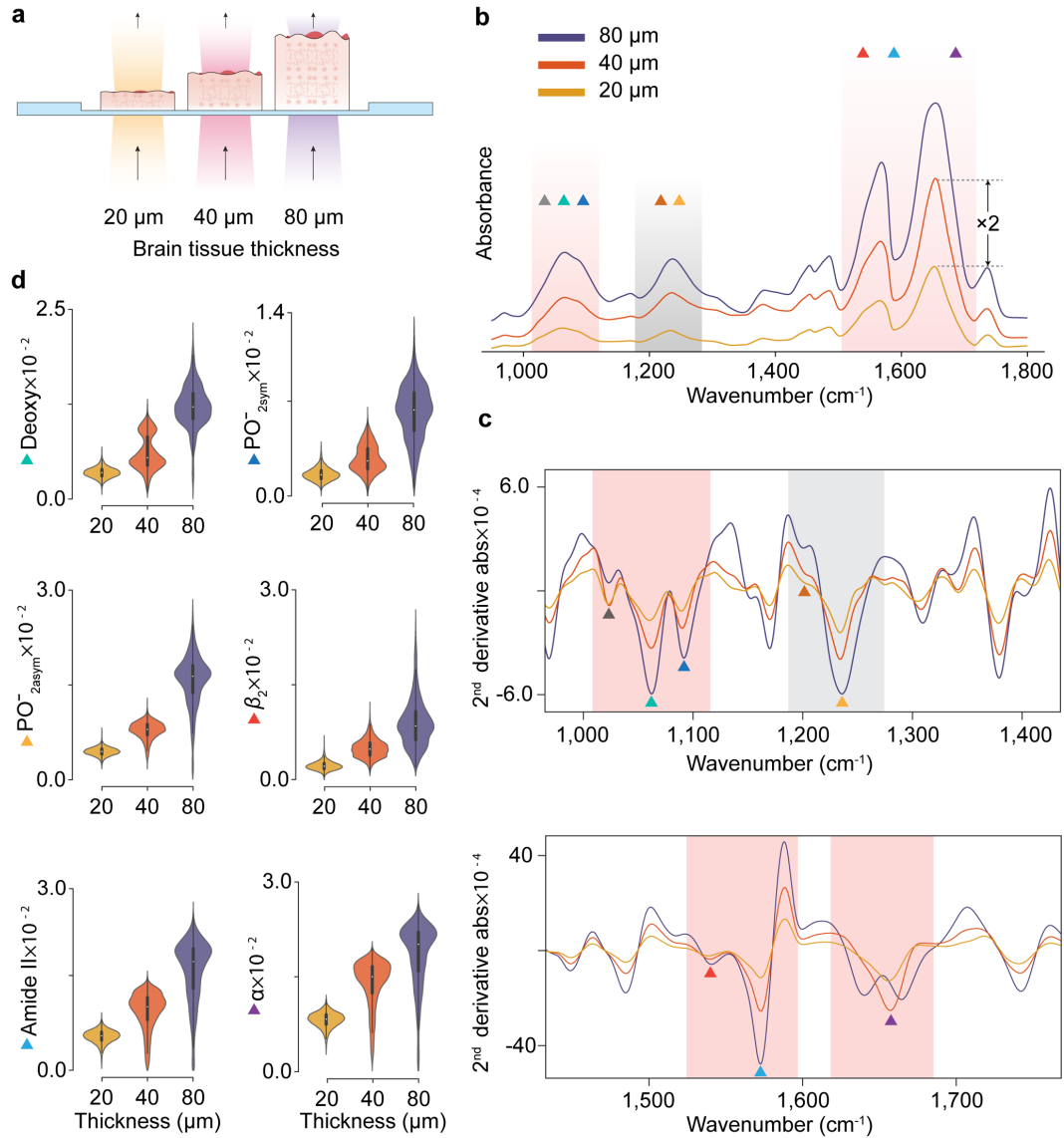

**Figure 2 | Tissue-thickness-dependent absorption spectra acquired by conventional mid-IR spectral imaging (MIRSI). a**, Sketch of three tissue sections with different thicknesses on standard IR-transparent substrates characterized in transmission mode. **b**, Measured absorption spectra from 80 μm, 40 μm, and 20 μm thick tissue sections. The absorption-band amplitudes consistently increase with tissue thickness. **c**, Second derivative of the absorption spectra for glycogen (1024 cm$^{-1}$ ▲), deoxyribose (1060 cm$^{-1}$ ▲), symmetric phosphate (1090 cm$^{-1}$ ▲), C-OH (1170 cm$^{-1}$ ▲), asymmetric phosphate (1236 cm$^{-1}$ ▲), β-sheet (1542 cm$^{-1}$ ▲), amide II (1574 cm$^{-1}$ ▲), and amide I (1660 cm$^{-1}$ ▲). **d**, Area integral distributions from $n_{pixels}$ = 12.5×10$^3$ different pixels along the second derivative absorption bands plotted for some of the above molecules. The individual pixels' integral value distributions from three different tissue thicknesses show quantitative and qualitative dependence on the spectral signatures.

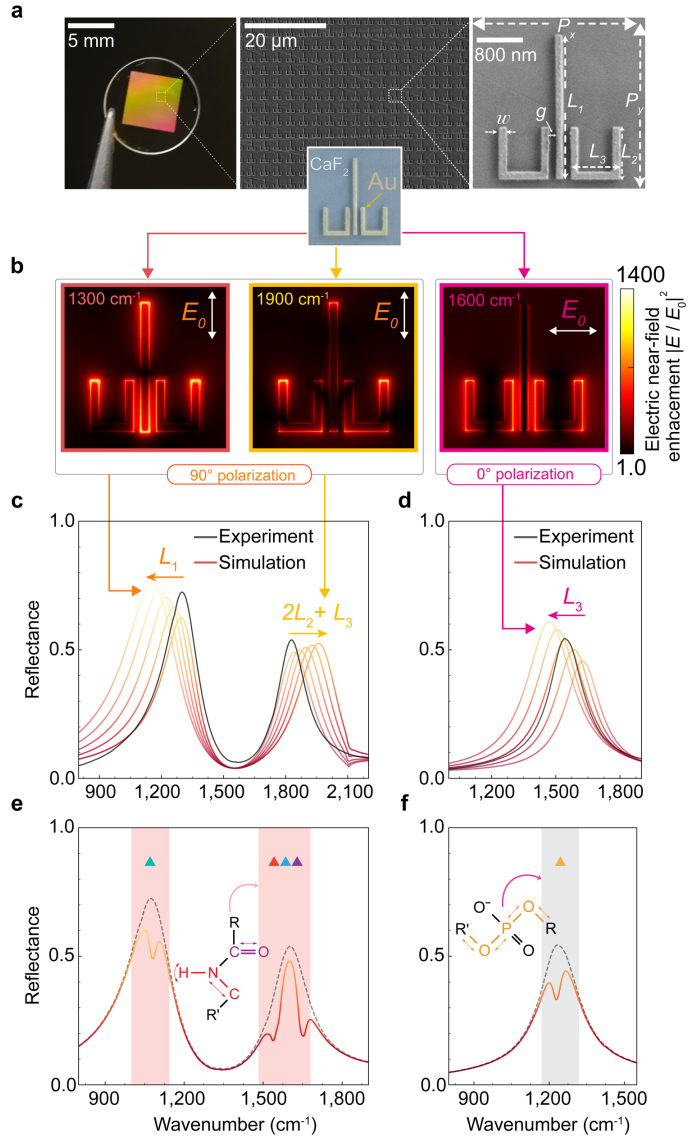

**Figure 3 | Polarization-multiplexed multi-resonant broadband plasmonic metasurface. a**, Photograph and SEM image of the two-dimensional array of plasmonic *ulu* nanostructures. The geometrical design parameters for the structure in the SEM image are $P_x$ = 3.2 μm, $P_y$ = 3.2 μm, $L_1$ = 2.4 μm, $L_2$ = 0.95 μm, $L_3$ = 0.8 μm, $g$ = 0.14 μm, and $w$ = 0.14 μm; 90 nm thick. **b**, Numerically simulated electric near-field intensity enhancement $|E/E_0|^2$ maps at different resonance wavenumbers for 0° and 90° polarizations. **c, d**, Experimentally measured (black curves) and simulated (red-yellow gradient) reflectance spectra of the polarization-multiplexed multi-resonant plasmonic metasurface. The three resonant peaks are tuned by adjusting the geometrical parameters of the *ulu* structures, such as $L_1$, $L_2$, and $L_3$, to spectrally match the absorption fingerprints of important biomolecules in the mid-Infrared region. **e, f**, Simulated reflectance spectra shows expected absorption signals associated with the molecular fingerprints of deoxyribose (1060 cm$^{-1}$ ▲), asymmetric phosphate (1236 cm$^{-1}$ ▲), β-sheet (1542 cm$^{-1}$ ▲), amide II (1574 cm$^{-1}$ ▲), amide I (1660 cm$^{-1}$ ▲). Molecular coupling with plasmonically enhanced

near-field hotspots at the resonance peaks (black dashed line) leads to dips in reflection spectra at wavenumbers corresponding to the vibrational resonances of the biomolecules (solid line).

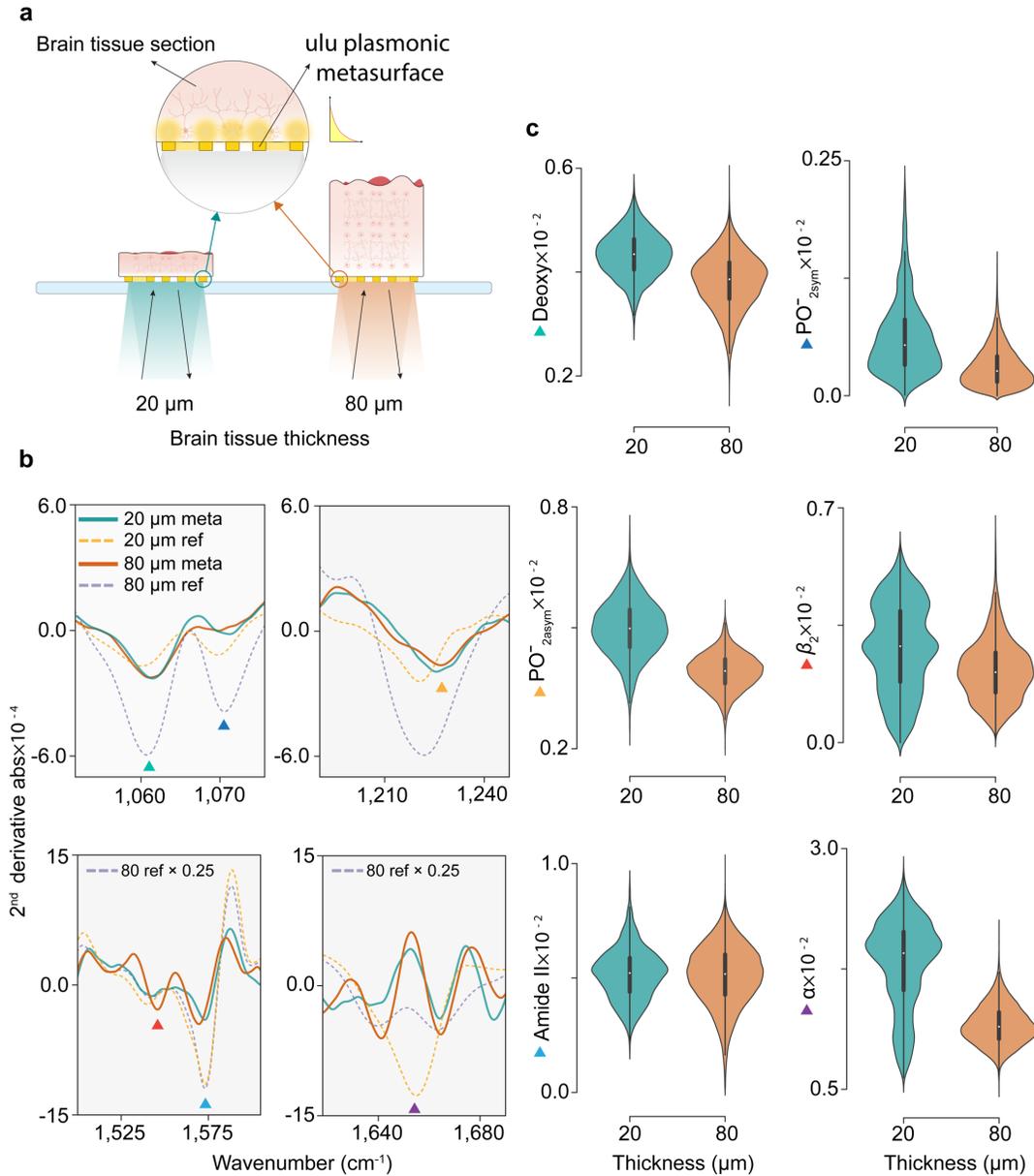

**Figure 4 | SE-MIRSI enables detection of tissue-thickness-independent molecular fingerprint.**
**a**, Sketch of two different brain tissue sections with thicknesses of 20 µm and 80 µm, mounted on metasurface patterned $CaF_2$ substrates. Mid-IR absorption spectra acquired in reflectance mode from the substrate side capture light-matter interactions from nanoscale electromagnetic hotspot volumes. **b**, Second derivative absorption spectra of 20 µm and 80 µm brain tissue sections collected by SE-MIRSI. Molecular absorption fingerprints of functional groups in brain tissue sections were detected via plasmonic metasurfaces (solid lines) and reference MIRSI absorption signals (dashed lines) were collected in transmission mode. Three different resonance peaks of the plasmonic metasurface were used to capture spectral signatures of deoxyribose (1060 cm$^{-1}$ ▲), symmetric phosphate (1090 cm$^{-1}$ ▲), asymmetric phosphate (1236 cm$^{-1}$ ▲), β-sheet (1542 cm$^{-1}$ ▲), amide II (1574 cm$^{-1}$ ▲), and amide I (1660 cm$^{-1}$ ▲). **c**, Area integral under second derivative absorption spectra distributions from $n_{pixels}$ = 12.5×10$^3$ different pixels for the six biomolecules mentioned above. Unlike in conventional MIRSI, the absorption signals do not increase with the tissue thickness in SE-MIRSI measurements.

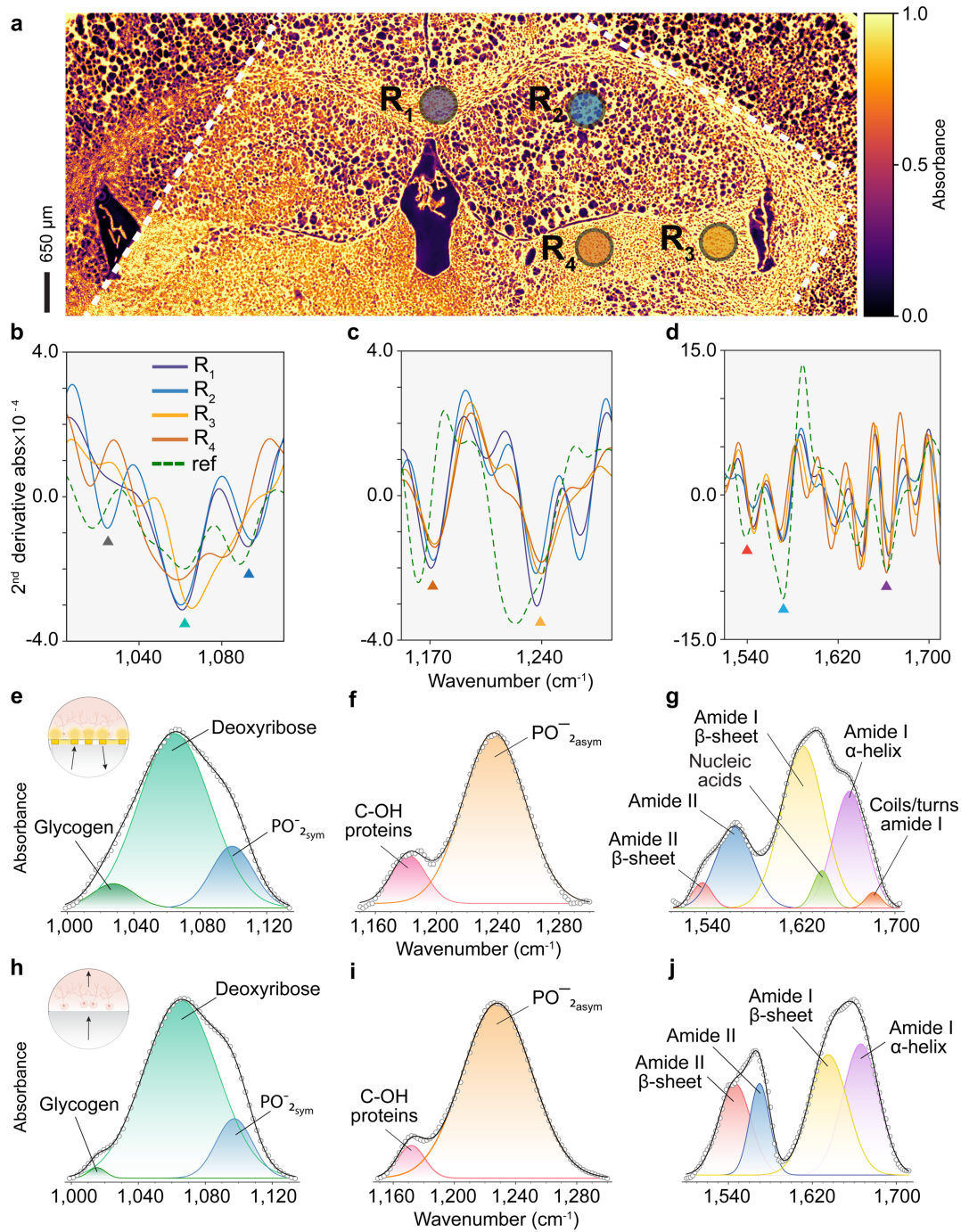

**Figure 5 | Metasurface-enabled molecular fingerprint retrieval from complex and spatially heterogeneous biological tissue regions. a**, Murine brain tissue section imaged in transmission mode at 1574 cm$^{-1}$ wavenumber showing areas both on and off the metasurface, which is enclosed by white dashed lines. The resonant plasmonic metasurface visibly enhances molecular absorbance when compared to tissue regions imaged on a standard substrate (regions outside of the dashed lines). **b, c, d**, Second derivative absorption spectra acquired in reflection mode from four different brain regions $R_1$, $R_2$, $R_3$, and $R_4$ on the metasurface (solid lines) and reference

spectrum (dashed line). Reference spectra (ref) of a brain tissue section with thickness 20 µm, mounted on a gold mirror, was referenced to a front-surface gold mirror. **e**, **f**, and **g**, Substrate band fitting analysis on three plasmonic resonance spectral regions for a brain tissue section on the metasurface confirm that the SE-MIRSI can capture fine spectral signatures from nanoscale surface volumes without the need for bulk tissue light-matter interactions. **h**, **i**, and **j**, Band fitting analysis on three different regions of absorption spectrum from tissue section on a standard MIR substrate.

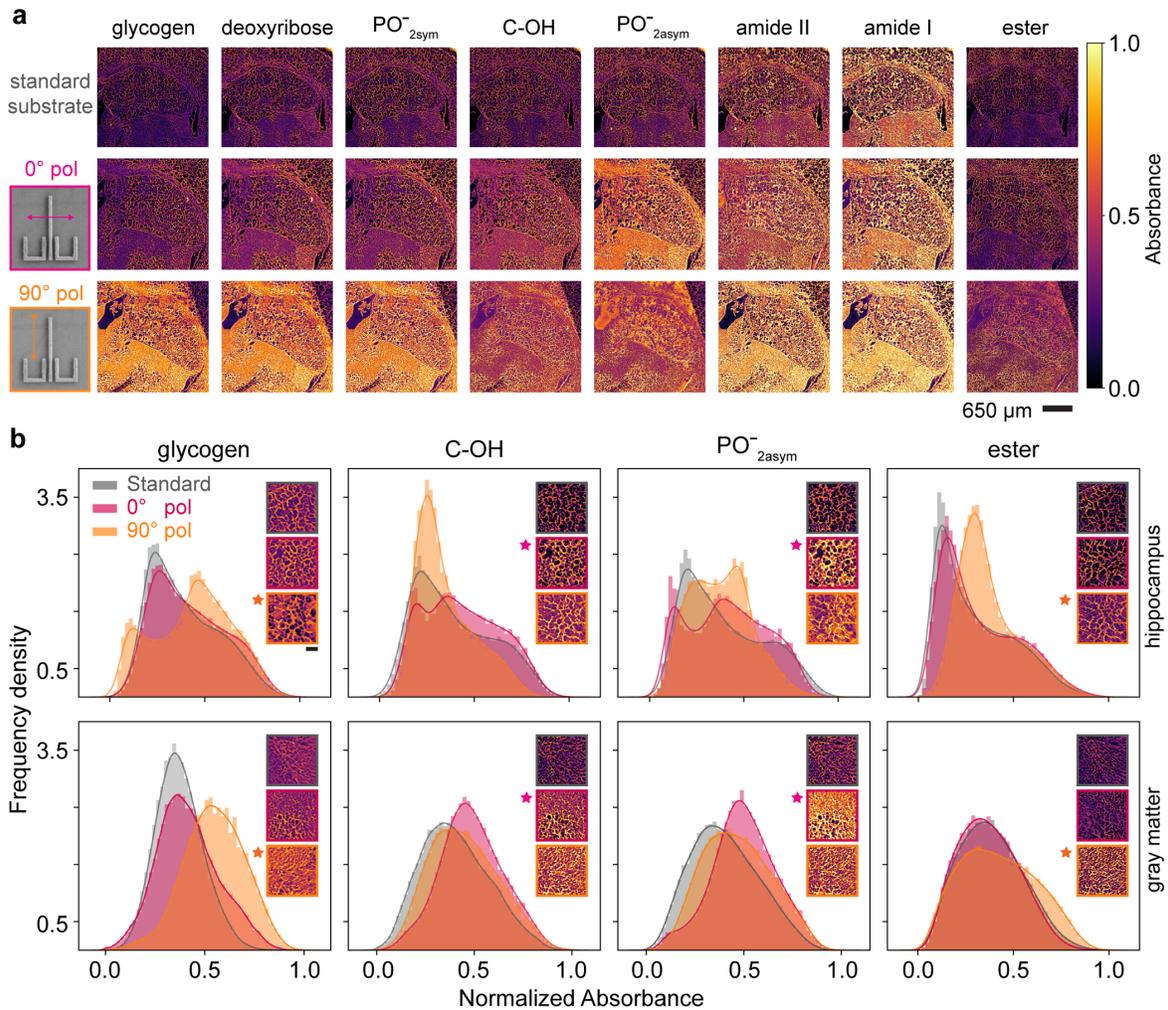

**Figure 6 | SE-MIRSI enhances molecular absorption resonances in morphologically heterogeneous tissue sections. a**, Two adjacent 20 μm-thick murine brain tissue sections are mounted on a standard CaF$_2$ substrate for MIRSI and a plasmonic metasurface for SE-MIRSI measurements. Infrared images at eight different absorption bands corresponding to various biomolecules show similar brain tissue regions measured on a standard substrate (top row), and on a metasurface illuminated at 0° (middle row) and 90° (bottom row) polarization modes. The plasmonic resonance excited with 0° polarization enhances absorption at C-OH and PO$^-_{2asym}$ bands and the two resonances excited with 90° polarization enhances glycogen, deoxyribose, PO$^-_{2sym}$, amide II, amide I, and ester bands. **b**, Histogram plots show absorbance frequency density distributions of 2.3×10$^5$ pixels measured by standard MIRSI and SE-MIRSI at orthogonal polarizations. Two different murine brain regions of interest from hippocampus (top row) and gray matter (bottom row) are shown for four molecules. Histogram distributions corresponding to plasmonic resonances denoted by stars show intensity shifts towards larger absorbance values confirming surface enhancement.

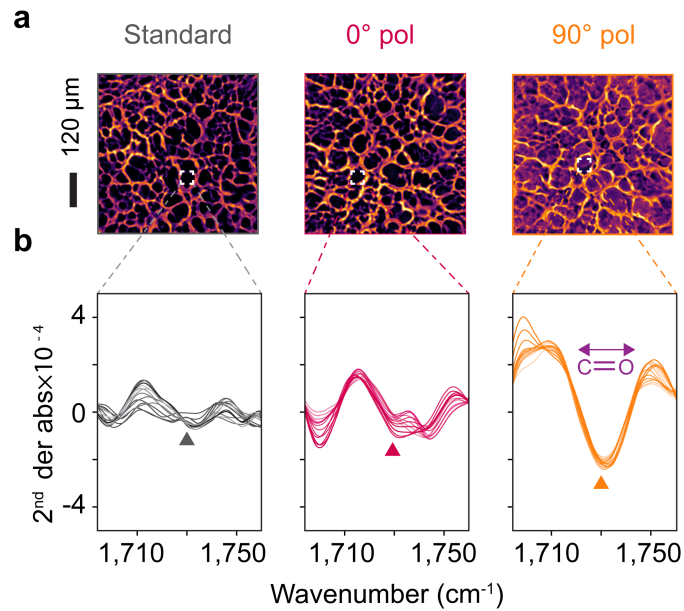

**Figure 7 | SE-MIRSI enabling chemical imaging of ultrathin regions in morphologically heterogeneous tissue sections. a**, Infrared images at 1730 cm$^{-1}$ corresponding to ester absorption band showing hippocampus region in brain tissue sections on a standard substrate and the metasurface at orthogonal polarizations. The plasmonic resonance at 90° polarization enhances optical absorption of lipid molecules, enabling their detection in the ultrathin tissue regions, which are not visible in standard MIRSI and off-resonance SE-MIRSI measurements. Ester absorption spectra from adjacent tissue regions, area enclosed by a white square. **b**, Characteristic ester absorption spectra from various pixels with high and low ester concentrations. SE-MIRSI at 90° polarization yields the highest ester absorbance due to the resonance enhancement spectrally aligned with the lipid band.